\documentclass{article}

\usepackage{microtype}
\usepackage{graphicx}
\usepackage{subfigure}
\usepackage{booktabs} % for professional tables
\usepackage{multirow}
\usepackage{hyperref}
\usepackage{makecell}
\usepackage{tabularx}
\usepackage{flushend}

\usepackage[accepted]{icml2023}

\usepackage{amsmath}
\usepackage{amssymb}
\usepackage{mathtools}
\usepackage{amsthm}

\usepackage[capitalize,noabbrev]{cleveref}

\theoremstyle{plain}

\theoremstyle{definition}

\theoremstyle{remark}

\newcommand{\model}{\texttt{MLSMM}~}
\usepackage[textsize=tiny]{todonotes}

\icmltitlerunning{MLSMM: Machine Learning Security Maturity Model}

\begin{document}

\twocolumn[
\icmltitle{MLSMM: Machine Learning Security Maturity Model}

% It is OKAY to include author information, even for blind
% submissions: the style file will automatically remove it for you
% unless you've provided the [accepted] option to the icml2023
% package.

% List of affiliations: The first argument should be a (short)
% identifier you will use later to specify author affiliations
% Academic affiliations should list Department, University, City, Region, Country
% Industry affiliations should list Company, City, Region, Country

% You can specify symbols, otherwise they are numbered in order.
% Ideally, you should not use this facility. Affiliations will be numbered
% in order of appearance and this is the preferred way.
\icmlsetsymbol{equal}{*}

\begin{icmlauthorlist}
\icmlauthor{Felix Viktor Jedrzejewski}{yyy}
\icmlauthor{Davide Fucci}{yyy}
\icmlauthor{Oleksandr Adamov}{yyy}

%\icmlauthor{}{sch}
%\icmlauthor{}{sch}
\end{icmlauthorlist}

\icmlaffiliation{yyy}{Department of Software Engineering, Blekinge Institute of Technology, Karlskrona, Sweden}

\icmlcorrespondingauthor{Felix Jedrzejewski}{felix.jedrzejewski@bth.se}

% You may provide any keywords that you
% find helpful for describing your paper; these are used to populate
% the "keywords" metadata in the PDF but will not be shown in the document
\icmlkeywords{Machine Learning, ICML}

\vskip 0.3in
]

% this must go after the closing bracket ] following \twocolumn[ ...

% This command actually creates the footnote in the first column
% listing the affiliations and the copyright notice.
% The command takes one argument, which is text to display at the start of the footnote.
% The \icmlEqualContribution command is standard text for equal contribution.
% Remove it (just {}) if you do not need this facility.

\printAffiliationsAndNotice{}  % leave blank if no need to mention equal contribution
%\printAffiliationsAndNotice{\icmlEqualContribution} % otherwise use the standard text.

\begin{abstract}
Assessing the maturity of security practices during the development of Machine Learning (ML) based software components has not gotten as much attention as traditional software development.
In this Blue Sky idea paper, we propose an initial Machine Learning Security Maturity Model (\model) which organizes security practices along the ML-development lifecycle and, for each, establishes three levels of maturity. 
We envision \model as a step towards closer collaboration between industry and academia.
\end{abstract}

\section{Introduction}
The release of ChatGPT and its fast popularity greatly contribute to the discussion about the role of AI in our society.
For example, several studies conducted on behalf of the German Federal Office for Information Security\footnote{\url{https://www.bsi.bund.de/EN/Service-Navi/Publikationen/Studien/Projekt_P464/Projekt_P464_node.html}} discuss the importance of regulations requiring industry to demonstrate their effort in addressing Machine Learning (ML) Security in their development practices.
% This calls for models to support the evaluation of ML-based system from a security perspective. 
The last decade of Adversarial Machine Learning (AML) research~\cite{biggio2018wild, cina2022wild} introduced multiple attacks and defense strategies.
While attackers seem to use publicly-available resources~\cite{tidjon2022threat}, multiple interviews with ML practitioners show that the industry is ill-prepared to handle potential attacks to its ML-based systems \cite{kumar2020adversarial} and reluctant to introduce security measures.
% which is observable solemnly after inspecting the title fragments of some publication in the last two years: ''I never thought about securing my Machine Learning System''\cite{boenisch2021never}, ''Security is not my field, I'm a stats guy''\cite{mink2023security}, ''Why do so?''\cite{grosse2023Machine} (Part of the first title submitted to arxiv). 
In a few instances, AML researchers have directly approached industry practitiones~\cite{boenisch2021never, mink2023security, grosse2023Machine}, conducted more realistic (e.g., in vivo) studies~\cite{apruzzese2022modeling}, and proposed actionable ML development models incorporating security measures \cite{zhang2022conceptualizing}.
In the \textit{traditional} software development community, the need for companies evaluating and incorporating security in their lifecycle has been addressed by several security maturity models~\cite{teodoro2011web, lipner2004trustworthy, weir2021infiltrating}. %BSIMM, Microsoft SDL
Nevertheless, besides the heavyweight ISO21827, there is a lack of ML security maturity models studied in academia and adopted in the industry.
% In general, most security maturity models require disproportional effort for companies to evaluate them \cite{rabii2020information}.
Based on the perspective regulations, attackers using AML, and the current low awareness about it in the industry based on empirical evaluations, we propose a lightweight domain-agnostic Machine Learning Security Maturity Model (\model).
The goals of the model are to i) evaluate the state-of-practice concerning the security of ML-development process within an organization, ii) support the organization in creating a roadmap to improve and prioritize their ML security stance in specific areas, and iii) increase ML security awareness across different teams (e.g., developers, architects, quality assurance).
\model is based on the established Security Assurance Maturity Model (SAMM\footnote{\url{https://owaspsamm.org/}}) proposed by the Open Worldwide Application Security Project (OWASP) and the Adversarial Threat Landscape for Artificial Intelligence Systems (ATLAS) taxonomy\footnote{\url{https://atlas.mitre.org/}} by MITRE. 
% The goal of MLSMM is to evaluate the ML security maturity level of a given company and pinpoint security gaps.
We foresee that \model will reduce the gaps between academia and industry fostering closer collaboration in which the first develops supportive tools and the latter provides real case scenarios, data, and study validation opportunities. 
Additionally, the roadmap \model can represent a starting point for compliance and certification procedures in the future. 

\begin{table*}[]
\centering
% \footnotesize
\caption{Excerpt of the proposed Machine-Learning Security Maturity Model.}
\label{tab:example-table}
\begin{tabularx}{\linewidth}{ccXXX}
\toprule
\textbf{ML Phase} & \textbf{Security Practice} & \textbf{Level 1} & \textbf{ Level 2} & \textbf{ Level 3} \\
\midrule
\multirow{2}{*}{Model Training}  & \makecell{Model Hardening\\ \tiny{\url{https://atlas.mitre.org/mitigations/AML.M0003/}}}& Model hardening is performed based on best-efforts (e.g., simple defense against model erosion) & Model hardening is standardized within the organization & Models are proactively hardened within the organization \\
\cline{2-5}
 & \makecell{Use Ensemble Methods\\\tiny{\url{https://atlas.mitre.org/mitigations/AML.M0006}}} & Simple ensemble models (e.g., voting) are introduced & Ensemble approaches are introduced or removed based on specific threats against the model & Ensembles are continuously shuffled to avoid leaking information to attackers \\
\bottomrule 
\end{tabularx}
\end{table*}

%\subsubsection{Software Engineering for ML}
%\subsubsection{Security Maturity Models}
OWASP SAMM is a state-of-practice maturity model facilitating the measurement, analysis, and improvement of software products' security and addressing essential stages in the software development process.
Its content is based on the experience and domain knowledge of industry security experts~\cite{brasoveanu2022security}.
% We will apply the the evaluation of maturity levels of mitigation techniques used in the respective stages of the ML development process.
Software engineering for ML follows a different development process than traditional software products.
Accordingly, for developing \model, we follow Amerishi et al.'s ML workflow~\cite{amershi2019software} consisting of nine stages being categorized as either model-oriented (i.e., model requirements, feature engineering, training, evaluation, deployment, and monitoring) or data-oriented (i.e., collection, cleaning, and labeling).
MITRE ATLAS is a taxonomy associating activities to mitigate attacks to real-world adversary tactics~\cite{zhang2023cross}. 
    % \begin{itemize}
    %     \item Nine stages of ML workflow being categorized as either data-oriented (e.g., collection, cleaning, and labeling) or model-oriented (e.g. , model requirements, feature engineering, training, evaluation, deployment, and monitoring) as a foundation for MLSMM \cite{amershi2019software}. This concepts describes the development process of ML models being model and domain agnostic.  
    %     \item Each Mitigations listed on \href{https://atlas.mitre.org/mitigations}{MITRE ATLAS} will be mapped on the ML workflow stages they occur. 
    %     \item MITRE ATLAS (Adversarial Threat Landscape for Artificial Intelligence Systems) is a knowledge base that lists defensive techniques linking the mitigated attacks and exemplary real-world case studies to them \cite{zhang2023cross}. 
    % \end{itemize}

%\cite{mitre-attck}

\section{MLSMM Prototype}
% explain why we follow a prescriptive approach (like SAMM) rather than a prescriptive one (like BSIMM)
% Besides the descriptive Building Security in Maturity Model (BSIMM), our model is prescriptive and built on the generative ML development process, suggesting and evaluating tested by MITRE mitigations in the respective ML development stage.
\model combines state-of-practice maturity evaluation techniques for software product security with state-of-the-art mitigation techniques assigned to particular stages within the ML development process.
Following OWASP SAMM, our proposed \model is prescriptive in nature---i.e., it provides high-level guidance and advices to an organization---rather than descriptive---i.e., providing a summary of what other organizations do.\footnote{For an example of descriptive security maturity model see BSIMM~\url{https://bsimm.com}}  
Table \ref{tab:example-table} presents an excerpt from the \textit{Model Training} phase of ML-components development.
The model is hierarchical; it starts with the nine phases of ML development~\cite{amershi2019software} each with a variable number of security practices from MITRE ATLAS associated with them.
Each security practice has three possible maturity levels where the activities on a lower level are typically easier to execute and require less formalization than the ones on a higher level.
At this initial stage, \model consists of 19 practices. 
A complete draft is available on the project website. \footnote{\url{https://anonymous.4open.science/r/MLSMM-EA81/}}
Similarly to SAMM, we propose a simple questionnaire measuring the maturity levels. We use ordinal-value answers to assess how well an organization fulfills the activities associated with a level. 
%Explanation of the table structure
% To evaluate the security maturity of company X applying an ML workflow, the model applicant iterates through the different ML development stages. For this example, the Model Training Phase is elaborated. 
% Based on the ML development documentation provided by company X, the model applicant inspects to what extent Model Hardening\footnote{\url{https://atlas.mitre.org/mitigations/AML.M0003/}} is performed during model training.
Based on the example in~\Cref{tab:example-table}, an organization reaches \textit{Maturity Level 1} in \textit{Model Hardening} once it performs activities such as adversarial training and network distillation.
However, these activities are performed ad-hoc and in an unstructured fashion~\footnote{A maturity level of zero indicates the complete lack of such activities}.
The organization reaches the next level once the answers to the questionnaire show evidence that hardening is a standardized practice for \textit{every} model.
The final level implies that model hardening is part of the model training process \textit{by design} rather than done in reaction to specific events.
In~\Cref{tab:example-table}, the next security practice assessed for \textit{Model Training} is \textit{Use Ensemble Methods}. 
The lowest maturity level indicates the presence of simple ensemble methods introduced during training without providing any security context.
Level 2 is reached once the use of ensemble methods is grounded in security activities identified before model development, such as threat modeling.
At Maturity Level 3, the organization continuously applies ensemble method shuffling to avoid information leakage. 
\model does not insist that an organization achieves the maximum maturity in every category as each organization should determine the target level, for each \textit{Security Practice}, that best fits their needs.

\section{Conclusion and Future Work}
We presented our idea for \model---an actionable, domain- and model-agnostic security maturity model to assess ML components developments based on existing industrial practices and procedures.
Our next steps are i) expand the model to cover additional ML security practices not included within MITRE ATLAS, ii) create a questionnaire to gather evidence to instantiate the model in practice, iii) validate the model and questionnaire with our industry partners regarding their usefulness and usability.

\bibliography{example_paper}
\bibliographystyle{icml2023}

% %%%%%%%%%%%%%%%%%%%%%%%%%%%%%%%%%%%%%%%%%%%%%%%%%%%%%%%%%%%%%%%%%%%%%%%%%%%%%%%
% %%%%%%%%%%%%%%%%%%%%%%%%%%%%%%%%%%%%%%%%%%%%%%%%%%%%%%%%%%%%%%%%%%%%%%%%%%%%%%%
% % APPENDIX
% %%%%%%%%%%%%%%%%%%%%%%%%%%%%%%%%%%%%%%%%%%%%%%%%%%%%%%%%%%%%%%%%%%%%%%%%%%%%%%%
% %%%%%%%%%%%%%%%%%%%%%%%%%%%%%%%%%%%%%%%%%%%%%%%%%%%%%%%%%%%%%%%%%%%%%%%%%%%%%%%
% \newpage
% \appendix
% \onecolumn
% \section{You \emph{can} have an appendix here.}

% You can have as much text here as you want. The main body must be at most $6$ pages long.
% For the final version, one more page can be added.
% If you want, you can use an appendix like this one, even using the one-column format.
% %%%%%%%%%%%%%%%%%%%%%%%%%%%%%%%%%%%%%%%%%%%%%%%%%%%%%%%%%%%%%%%%%%%%%%%%%%%%%%%
% %%%%%%%%%%%%%%%%%%%%%%%%%%%%%%%%%%%%%%%%%%%%%%%%%%%%%%%%%%%%%%%%%%%%%%%%%%%%%%%

\end{document}